\documentclass{iopart}
\usepackage{iopams}
\usepackage{amssymb}
\usepackage{bm}
\usepackage{graphicx}
\begin{document}
\title[Quantum mechanics of photons]{Quantum-mechanical description of optical beams}
\author{Iwo Bialynicki-Birula}
\address{Center for Theoretical Physics, Polish Academy of Sciences\\
Aleja Lotnik\'ow 32/46, 02-668 Warsaw, Poland}
\ead{birula@cft.edu.pl}
\author{Zofia Bialynicka-Birula}
\address{Institute of Physics, Polish Academy of Sciences\\
Aleja Lotnik\'ow 32/46, 02-668 Warsaw, Poland}

\begin{abstract}

Quantum mechanics of photons is derived from the theory of representations of the Poincar\'e group developed by Wigner. This theory places helicity as the most fundamental property; angular momentum and polarization are secondary characteristics. The properties of the beams of light are shown to be fully determined by the quantum states of the photons. Polarization of light beams is explained as the freedom to chose an arbitrary combination of the helicity states. Quantum mechanics of photons enables one to give a precise meaning to the concept of wave-particle duality.
\end{abstract}

\submitto{Journal of Optics}

\section{Introduction}

This paper offers a complementary viewpoint to the one presented in recent reviews [1--4] devoted to the angular momentum of optical beams. Optical beams are made of photons and the wave-particle duality helps to make the properties of optical beams more transparent. In particular, the concept of the angular momentum when applied to photons acquires some properties that are hard to see in the solutions of Maxwell equations.

Photon is an elementary particle---a quantum of the electromagnetic field---and the particle point of view makes some properties of the optical beams unambiguous. We believe that by taking the quantum mechanics of photons as the basis, one is able to further elucidate the issues discussed in [1--4]. We shall show that by associating photon quantum numbers with the corresponding beams of electromagnetic radiation we gain a new insight into the physical properties of these beams. Some of the aspects of this quantum-classical correspondence were touched upon in our previous papers [5--7] but here we do it in a systematic way.

There are three intertwined physical properties that appear in the description of beams of electromagnetic radiation endowed with angular momentum. These are: {\em helicity, polarization and spin}. Quite often the researchers attached different meanings to these concepts and that led to various misunderstandings. An additional source of confusion is the use of various approximations that give results that contradict those predicted by the exact theory. We believe that the fundamental issues should be addressed within the exact theory and in this work we always use the full relativistic theory. Placing the particle concept at the heart of our analysis, we gain a universal platform that helps to disentangle various seemingly contradictory ways of beam description. Of the three properties listed above, the most rudimentary notion for massless particles is that of {\em helicity}.

A significant advantage of using the quantum mechanics of photons as the basis is the possibility of applying a mathematically impeccable tool: the theory of representations of the Poincar\'e group. The quantum-mechanical viewpoint suggests also a systematic use of quantum numbers for photon states as a convenient characterization of the optical beams. Quantum numbers are very often associated directly with the symmetries; they are just eigenvalues of the generators of the symmetries and these operators are unambiguously defined.

In particle physics helicity is defined as the projection of the total angular momentum on the direction of propagation, measured in units of $\hbar$. This definition applies to massive and massless particles but it plays a different role in these two cases. In the massive case, helicity is used to label {\em different states} of a particle within a given representation of the Poincar\'e group. In this case for a particle of spin $s$ there are $2s+1$ different helicity states that form the basis of a single representation. In contrast, for massless particles helicity takes on only two extreme values, $s$ and $-s$, that label {\em two different representations} of the Poincar\'e group. For massless particles helicity is an invariant; it is not changed by a change of the reference frame.

There is a simple physical explanation why the number of different helicity states is $2s+1$ for massive and 2 for massless particles. Massive particles can be stopped (i.e. they can be described in their rest frame) and all $2s+1$ spin states are then accessible while massless particles cannot be stopped; in all reference frames they move with the speed of light. Of course, we cannot be certain that the rest mass of photons is exactly zero but all different measurements \cite{gn,betal} give an incredibly small upper limit (less than $10^{-50}$kG) on the photon mass. Therefore, it is practically impossible to distinguish in laboratory experiments between photons with such a tiny rest mass and massless photons. Massive photons would indeed be markedly different because they would have three instead of two helicity states. However, it has been shown \cite{bg} that in the limit, when the mass of such a photon mass goes to zero, the third state decouples smoothly from the electrically charged matter. Thus, the consideration of the three photon helicity states, like in \cite{wxq}, would have no observable consequences. The crucial difference between massive and massless particles affects also the related notions of spin and polarization.

Of course, the analysis based on the Poincar\'e group applies only to photons moving in empty space. When photons interact with matter, helicity is no longer conserved (cf. Sec.~\ref{med}).

\section{Description of photon states in the momentum representation}\label{mom}

The continuous Poincar\'e group consist of three space translations, one time translation, three rotations, and three transformations to a moving frame (boosts). There are two interpretations of the transformations: passive and active. We shall stick here to the active interpretation because, in our opinion, it more forcefully underscores the role of generators.

Before introducing a more intricate description of photon wave mechanics in the position representation, we give the description of photon states in the momentum representation. This description can be based on a solid foundation constructed by Wigner \cite{wig} and by Bargmann and Wigner \cite{barwig}. The main result of their work is a full classification of the unitary representations of the Poincar\'e group. We are here interested in the two representations that pertain to photons: one for photons with the left-handed circular polarization (positive helicity) and second for photons with the right-handed polarization (negative helicity). These two representations operate separately in the space of complex-valued wave functions $f_+(\bm{k})$ and $f_-(\bm{k})$. Thus, the theory of representations gives an unambiguous definition of helicity for massless particles; the value of helicity distinguishes between two different representations of the Poincar\'e group.

Under the continuous Poincar\'e transformations the wave functions $f_+(\bm{k})$ and $f_-(\bm{k})$ transform separately. Unitary character of the representations means that the transformation of the Poincar\'e group, apart from the obvious change of the argument ${\bm k}$, results only in the change of the phase of the wave functions. Under space and time translations they transform in the same fashion,
\begin{eqnarray}\label{tr}
f'_\pm({\bm k})=\rme^{\rmi({\bm k}\cdot{\bm r}-\omega t)}f_\pm({\bm k}).
\end{eqnarray}
We adopted in this formula the sign convection of standard quantum mechanics. Namely, the wave functions with the time dependence $\rme^{-\rmi\omega t}$ describe states with positive energy.

Under the Lorentz transformation the two representations pick up the phase factors differing only by the {\em opposite} sign of the phase \cite{wig},
\begin{eqnarray}\label{lorentz}
f_\pm({\bm k}')=\rme^{\pm\rmi\Theta}f_\pm({\bm k}),
\end{eqnarray}
where ${\bm k}'$ is the Lorentz-transformed wave vector. This is why it is not allowed to take simply a sum of the left-handed and right-handed wave functions. The situation is analogous to that of massive spinning particles; we cannot add the wave functions describing an electron with spin up and spin down but we must combine them into a two-dimensional Pauli spinor.

The two representations do not mix only under the continuous transformations of the Poincar\'e group; under the discrete transformations (space and time reflections) they are interchanged,
\begin{eqnarray}\label{pt}
f'_\pm(-\bm{k})=f_\mp(\bm{k})\qquad{\rm space\;reflection},\\
f'_\pm(-\bm{k})=f^*_\mp(\bm{k})\qquad{\rm time\;reflection}.
\end{eqnarray}

General one-photon states of arbitrary polarization are described by a pair of functions that may be conveniently combined into a two-dimensional vector,
\begin{eqnarray}\label{pwf}
{\bm{\mathfrak f}}({\bm k})=\left(\begin{array}{c}f_+({\bm k})\\f_-({\bm k})\end{array}\right).
\end{eqnarray}
The Hilbert space of these functions is endowed with the scalar product:
\begin{eqnarray}\label{sp}
\langle{\bm{\mathfrak f}}|{\bm{\mathfrak g}}\rangle=\sum_\lambda\int\!\frac{d^3k}{k}\,
f_\lambda^*(\bm{k})g_\lambda(\bm{k}),\qquad \lambda=(+,-),
\end{eqnarray}
where $k=|\bm{k}|$. This scalar product is relativistically invariant since the wave functions change only their phases and the factor $1/k$ makes the volume element on the light-cone invariant.

In what follows, we will need the matrices that act in the two-dimensional space of the two helicity components. Analogous matrices, acting on the upper and lower components of the bispinor, were introduced by Dirac \cite{dir} and denoted by $\{\rho_1,\rho_2,\rho_3\}$. They are numerically equal to the Pauli matrices but they do not act on the components of spinors. The matrix $\rho_3$ will be singled out and denoted by ${\hat{\lambda}}$ since it plays a special role. It may be called the helicity operator in the momentum representation because acting on ${\bm{\mathfrak f}(\bm{k})}$ it gives the sign of helicity $\lambda$,
\begin{eqnarray}\label{hm}
{\hat{\lambda}}{\bm{\mathfrak f}}({\bm k})=\left(\begin{array}{c}f_+({\bm k})\\-f_-({\bm k})\end{array}\right).
\end{eqnarray}

To allow for the probabilistic interpretation of the wave functions (\ref{pwf}) we must impose the normalization condition:
\begin{eqnarray}\label{norm}
\int\!\frac{d^3k}{k}|{\bm{\mathfrak f}(\bm{k})}|^2=\sum_\lambda\int\!\frac{d^3k}{k}
|f_\lambda(\bm{k})|^2=1.
\end{eqnarray}

The most important quantum numbers characterizing the states of the photon are the eigenvalues of the ten generators of the Poincar\'e group: the energy $\hat{H}$ (generator of time translations), the momentum ${\hat{\bm P}}$ (generator of space translations), the angular momentum ${\hat{\bm J}}$ (generator of rotations), and ${\hat{\bm N}}$ (generator of boosts). The generators acting on the photon wave functions in the momentum representation have the form (cf., for example, \cite{qed,oam})
\numparts
\begin{eqnarray}
{\hat H}=\hbar\omega_{\bm k},\label{gena}\\
{\hat{\bm P}}=\hbar{\bm k},\label{genb}\\
{\hat{\bm J}}=-\rmi\hbar{\bm k}\times{\bm D}+\hbar{\hat\lambda}{\bm n}_{\bm k},\label{genc}\\
{\hat{\bm N}}={\rmi}\hbar\omega_{\bm k}{\bm D},\label{gend}
\end{eqnarray}
\endnumparts
where ${\bm n}_{\bm k}={\bm k}/|{\bm k}|$, and ${\bm D}$ stands for the covariant derivative on the light cone,
\begin{eqnarray}\label{cd}
{\bm D}={\bm{\partial}}-\rmi{\hat\lambda}{\bm\alpha}({\bm k}),
\end{eqnarray}
where ${\bm{\partial}}$ denotes the gradient in momentum space ${\bm{\partial}}=\{\partial/\partial_{k_x},\partial/\partial_{k_y},
\partial/\partial_{k_z}\}$. The vector ${\bm\alpha}({\bm k})$ is an analogue of the potential; it depends on the (arbitrary) phase of the wave function. However, its curl ${\bm{\partial}}\times{\bm\alpha}(\bm{k})=-{\bm k}/k^3$ (the analogue of the curvature) is unique. The origin of the covariant derivative ${\bm D}$ is explained in Sec.\ref{conn} where the formulas (\ref{gena}-\ref{gend}) for the generators are derived. The generators (\ref{gena}-\ref{gend}) are Hermitian with respect to the scalar product (\ref{sp}). Note that the formulas (\ref{genb}) and (\ref{genc}) confirm our interpretation of ${\hat\lambda}$ as the helicity operator because the projection of ${\hat{\bm J}}$ on the direction of ${\hat{\bm P}}$ is equal to $\hbar{\hat\lambda}$.

\section{Description of photon states in the position representation}\label{st}

In free space the formulation of the quantum mechanics of photons in the momentum representation is complete but it is not sufficient to describe the interaction of photons with matter since matter is distributed in position space and this requires the position representation\footnote{The term position representation seems to be more adequate than often used {\em coordinate} representation}. In nonrelativistic quantum mechanics the transition between the momentum representation and the position representation involves only a straightforward Fourier transformation. This simple method does not work for photons because, as has been observed already by Pauli \cite{pauli}, it violates the principle of {\em locality}. This principle in this context may be formulated as follows: the value of the photon wave function in space at the point $\bm r'$ after any transformation $\mathcal P$ of the Poincar\'e group may depend {\em only} on its value at the point $\bm r$ which was mapped into the point $\bm r'$ by the transformation $\mathcal P$.

The proper form of the photon wave functions ${\bm\Psi_\pm({\bm r},t)}$ of both helicities in the position representation may be obtained taking into account the following requirements:
\begin{enumerate}
\item{${\bm\Psi_\pm({\bm r},t)}$ must be a local field forming a representation of the Poincar\'e group.}
\item{${\bm\Psi_\pm({\bm r},t)}$ must be directly related to the electromagnetic fields obeying the Maxwell equations.}
\item{${\bm\Psi_\pm({\bm r},t)}$ in its Fourier decomposition may contain only positive energies (frequencies).}
\end{enumerate}
All these requirements are met by the following definitions:
\begin{eqnarray}\label{def}
{\bm\Psi}_+(\bm r,t)={\bm F}^{(+)}(\bm r,t),\quad {\bm\Psi}_-(\bm r,t)={\bm F}^{*(+)}(\bm r,t),
\end{eqnarray}
where ${\bm F}(\bm r,t)$ is the Riemann-Silberstein (RS) vector and $(+)$ denotes the part containing only positive frequencies. The RS vector is a very convenient representation of the electromagnetic field. It is a complex vector field built from the electric and magnetic fields as follows \cite{bb,pwf}:
\begin{eqnarray}\label{rs}
{\bm F}(\bm r,t)= \frac{\bm D(\bm r,t)}{\sqrt{2\varepsilon_0}}+\rmi\frac{\bm B(\bm r,t)}{\sqrt{2\mu_0}}.
\end{eqnarray}
The Maxwell equations satisfied by ${\bm D(\bm r,t)}$ and ${\bm B(\bm r,t)}$ imply that the RS vector satisfies the equations:
\begin{eqnarray}\label{max}
\rmi\partial_t{\bm F}(\bm r,t)
=c {\bm\nabla}\times{\bm F}(\bm r,t).
\end{eqnarray}
Every vector field ${\bm F}(\bm r,t)$ satisfying these equations can be expanded into plane waves as follows:
\begin{eqnarray}\label{pwdec}
\qquad\fl{\bm F}(\bm r,t)=\sqrt{\hbar c}\int\!\frac{d^3k}{(2\pi)^{3/2}}{\bm e}(\bm k)\left[f_+(\bm k)\rme^{\rmi(\bm k\cdot\bm r-\omega t)}+f^*_-(\bm k)\rme^{-\rmi(\bm k\cdot\bm r-\omega t)}\right],
\end{eqnarray}
where the factor $\sqrt{\hbar c}$ is needed for dimensional reasons and the normalized vector ${\bm e}(\bm k)$ satisfies the equation
\begin{eqnarray}\label{algr}
c{\bm k}\times{\bm e}(\bm k)=-\rmi\,\omega\,{\bm e}(\bm k).
\end{eqnarray}
The vector ${\bm e}(\bm k)$ coincides with the standard polarization vector of classical electromagnetism for circular polarization (p.~299 in \cite{jack}) and it can be chosen as \cite{bb0,bb}:
\begin{eqnarray}\label{pol}
\fl\quad{\bm e}({\bm k})=\frac{1}{\sqrt{2}\,k\,\sqrt{k_x^2+k_y^2}}
\!\left[\begin{array}{c}\vspace{0.2cm}
-k_xk_z+\rmi k k_y\\\vspace{0.2cm}
-k_yk_z-\rmi k k_x\\
k_x^2+k_y^2\end{array}
\right]=\frac{1}{\sqrt{2}}\left[\begin{array}{c}
-\cos\theta\cos\phi+\rmi\sin\phi\\
-\cos\theta\sin\phi-\rmi\cos\phi\\
\sin\theta
\end{array}\right],
\end{eqnarray}
where $\phi$ and $\theta$ are the directional angles of $\bm k$.

By comparing the formula (\ref{pwdec}) with the postulated definitions of the photon wave functions (\ref{def}), we obtain the following expansions into plane waves of the photon wave functions in the position representation:
\numparts
\begin{eqnarray}
{\bm\Psi}_+(\bm r,t)=\sqrt{\hbar c}\int\!\frac{d^3k}{(2\pi)^{3/2}}{\bm e}(\bm k)f_+(\bm k)\rme^{\rmi(\bm k\cdot\bm r-\omega t)},\label{rwf}\\
{\bm\Psi}_-(\bm r,t)=\sqrt{\hbar c}\int\!\frac{d^3k}{(2\pi)^{3/2}}{\bm e}^*(\bm k)f_-(\bm k)\rme^{\rmi(\bm k\cdot\bm r-\omega t)}.\label{lwf}
\end{eqnarray}
\endnumparts

The general state of a photon may have components with both helicities so that it is described by a two-component vector,
\begin{eqnarray}\label{col}
{\underline{\bm{\Psi}}}({\bm r},t)=\left(\begin{array}{c}{\bm{\Psi}}_+({\bm r},t)\\{\bm{\Psi}}_-({\bm r},t)\end{array}\right).
\end{eqnarray}

The wave equation obeyed by this function follows from the equation (\ref{max}) for the RS vector,
\begin{eqnarray}\label{curl}
\rmi\partial_t{\underline{\bm{\Psi}}}({\bm r},t)=c{\hat{\lambda}} {\bm\nabla}\times{\underline{\bm{\Psi}}}({\bm r},t).
\end{eqnarray}
The meaning of the helicity matrix $\hat{\lambda}$ in these formula is the same as in the momentum representation, that is $\hat{\lambda}$ is a diagonal $6\times 6$ matrix with the components $\hat{\lambda}={\rm diag}\{1,1,1,-1-1-1\}$. Since the right-hand side in (\ref{curl}) is divergence-free, the wave functions satisfying this equation must be divergence-free (apart from some constant part representing static fields),
\begin{eqnarray}\label{div}
{\bm\nabla}\!\cdot\!{\bm\Psi}_\pm({\bm r},t)=0.
\end{eqnarray}
This requirement restricts the set of allowed operators acting on $\bm{\underline{\bm{\Psi}}}({\bm r},t)$ to those that respect (\ref{div}). In particular, the naive position operator for photons---multiplication by $\bm r$---is excluded.

The equations (\ref{curl}) can also be given a form similar to the Dirac equation \cite{pwf0,pwf}:
\begin{eqnarray}\label{pwe}
\partial_t{\underline{\bm{\Psi}}}({\bm r},t)=c{\hat{\lambda}}\left({\bm s}\!\cdot\!{\bm\nabla}\right){\underline{\bm{\Psi}}}({\bm r},t)
\end{eqnarray}
where ${\bm s}$ are the spin 1 matrices,
\begin{eqnarray}\label{spin}
{\bm s}=\left\{\left[\begin{array}{ccc}
0&0&0\\
0&0&-\rmi\\
0&\rmi&0
\end{array}\right],
\left[\begin{array}{ccc}
0&0&\rmi\\
0&0&0\\
-\rmi&0&0
\end{array}\right],
\left[\begin{array}{ccc}
0&-\rmi&0\\
\rmi&0&0\\
0&0&0
\end{array}\right]\right\}.
\end{eqnarray}

We have tentatively identified the photon wave functions in the momentum representation with the coefficients in the Fourier expansion (\ref{pwdec}). There are two different arguments that justify this choice.

The first argument is based on the fact that the expansion coefficients $f_\pm(\bm k)$ have the same transformation properties as the representations of the Poincar\'e group (\ref{tr}) and (\ref{lorentz}) appropriate for photons. This follows from the transformation properties of the RS vector, i.e. from the transformation properties of the electromagnetic fields (see, for example, p. 558 of \cite{jack}). Considering infinitesimal transformations, we find the following form of the ten generators of the Poincar\'e group in the position representation:
\numparts
\begin{eqnarray}
{\hat{\bm P}}=-\rmi\hbar{\bm\nabla},\label{gena1}\\
{\hat H}=-i\hbar{\hat\lambda}{\bm s}\!\cdot\!{\bm\nabla},\label{genb1}\\
{\hat{\bm J}}=-\rmi\hbar{\bm r}\times{\bm\nabla}+\hbar{\bm s},\label{genc1}\\
{\hat{\bm N}}=\rmi\hbar({\bm r}\partial_t\times+c^2t{\bm\nabla})
+c{\hat\lambda}\hbar{\bm s}.\label{gend1}
\end{eqnarray}
\endnumparts

By applying the operators (\ref{gena1}-\ref{gend1}) to the photon wave functions (\ref{rwf}-\ref{lwf}), after performing all operations under the integral sign and integrating by parts, we obtain the expressions (\ref{gena}-\ref{gend}) for the generators in the momentum representation. This proves the consistency of the descriptions of photons in the momentum representation and in the position representation.

As a byproduct of these calculations, we find that the vector $\bm\alpha$ appearing Eqs.~(\ref{gena}-\ref{gend}) is to be identified with:
\begin{eqnarray}\label{alpha}
\bm\alpha(\bm k)=\rmi\{{\bm e}^*(\bm k)\!\cdot\!\partial_x{\bm e}(\bm k),{\bm e}^*(\bm k)\!\cdot\!\partial_y{\bm e}(\bm k),{\bm e}^*(\bm k)\!\cdot\!\partial_z{\bm e}(\bm k)\}.
\end{eqnarray}
Our choice of $\bm e$ is most convenient in the description of beams moving in the $z$-direction because in this case we obtain:
\begin{eqnarray}\label{alpha1}
\bm\alpha(\bm k)=\frac{\cot\theta}{k}\{-\sin\phi,\cos\phi,0\},
\end{eqnarray}
and the $z$-component of the total angular momentum in the momentum representation reduces to $J_z=-\rmi\hbar(k_x\partial_y-k_y\partial_x)$ because the helicity term $\hbar{\hat\lambda}{\bm n}_{\bm k}$ in (\ref{genc}) is canceled by the contribution from $\bm\alpha$ in $\bm D$.

The second argument is based on the expressions for the energy and momentum of the photons. The expectation values of the energy and momentum operators given by (\ref{gena}) and (\ref{genb}) are:
\begin{eqnarray}\label{exp}
\quad\fl\langle H\rangle=\sum_\lambda\int\!\frac{d^3k}{k}f_\lambda^*(\bm k)\hbar\omega_{\bm k}f_\lambda(\bm k),\quad\langle{\bm P}\rangle=\sum_\lambda\int\!\frac{d^3k}{k}f_\lambda^*(\bm k)\hbar{\bm k}f_\lambda(\bm k).
\end{eqnarray}
The same expressions are obtained when the energy $E$ and momentum $\bm P$ of the electromagnetic field is expressed in terms of the RS vector,
\begin{eqnarray}\label{enmom}
\quad\fl E&=\int\!d^3r\,\left[\frac{{\bm D}(\bm r,t)\!\cdot\!{\bm D}(\bm r,t)}{2\varepsilon_0}+\frac{{\bm B}(\bm r,t)\!\cdot\!{\bm B}(\bm r,t)}{2\mu_0}\right]=\int\!d^3r\,{\bm F}^*(\bm r,t)\!\cdot\!{\bm F}(\bm r,t),\\
\quad\fl {\bm P}&=c\int\!d^3r\,{\bm D}(\bm r,t)\times{\bm B}(\bm r,t)=-\rmi\int\!d^3r\,{\bm F}^*(\bm r,t)\times{\bm F}(\bm r,t),
\end{eqnarray}
and these integrals, in turn, are evaluated with the use of the Fourier expansion (\ref{pwdec}). This calculation also justifies the presence of the factors $\sqrt{\hbar c}$ in the Fourier expansions of the RS vector (\ref{pwdec}) and the photon wave functions (\ref{rwf}) and (\ref{lwf}).

The photon wave functions in the position representation do not have all the properties of the wave functions of massive particles and some physicists even denied their existence. Edwin Power wrote \cite{ep}: ``Strictly speaking there are no such functions!''. These controversies arose due to the absence of the standard probabilistic interpretation of the wave functions $\Psi_\pm$ since there is no local photon density.

The modulus squared of the wave function $|\Psi_\pm|^2$ is a local quantity. However, it does not have the dimension of the probability density but of the energy density (exactly like ${\bm F}^*\!\cdot\!{\bm F}$). The right dimensionality is restored after the division by the total energy. The dimensionless quantity $p_\pm(V)$,
\begin{eqnarray}\label{pv}
p_\pm(V)=\frac{\int_Vd^3r|\Psi_\pm({\bm r},t)|^2}{\int d^3r\left[|\Psi_+({\bm r},t)|^2+|\Psi_-({\bm r},t)|^2\right]},
\end{eqnarray}
is to be interpreted as the expected fraction of the total photon energy (with the helicity $\pm 1$) to be found at the time $t$ in the volume $V$. To put it succinctly: the photon is where its energy is.

Landau and Peierls \cite{lp,good} proposed to define the photon wave function $\bm{\Phi}$ in the position representation with the correct dimension. However, this definition requires the division by $\sqrt{\hbar\omega}$ which is a nonlocal operation in the position space,
\begin{eqnarray}\label{lp}
\bm{\Phi}({\bm r},t)=``\frac{1}{\sqrt{\hbar\omega}}"{\bm\Psi}({\bm r},t)=\frac{\pi}{\hbar c}\int\frac{d^3r'}{(2\pi|\bm r-\bm r'|)^{5/2}}{\bm\Psi}({\bm r}',t).
\end{eqnarray}
Of course, there are two Landau-Peierls wave functions for the two helicities. Despite their nonlocal character, these functions have some interesting applications. In particular, the average values of all the observables (\ref{gena1}-\ref{gend1}) evaluated according to the quantum-mechanical prescription in the position representation,
\begin{eqnarray}\label{av}
\langle{G}\rangle=\int\!d^3r\bm{\Phi}^*({\bm r},t){\hat G}\bm{\Phi}({\bm r},t).
\end{eqnarray}
coincide with the average values evaluated in the momentum representation.

\section{Wave-particle duality}\label{conn}

The purpose of this section is to give the precise meaning to the concept of the wave-particle duality. We may phrased it as follows: there is one-to-one correspondence between the wave function describing the quantum state of the photon and the classical solution of the Maxwell equations. This correspondence holds in two contexts: in the quantum mechanics of photons and in the full quantized theory of the the electromagnetic field.

In the first case, the mathematical expression of this correspondence is the Fourier expansion of the RS vector (\ref{pwdec}). In this formula we have the combination of the electric and magnetic fields and on the right hand side under the integral we have the photon wave functions of both helicities.

In the second case, this correspondence is  expressed in terms of the coherent states of the electromagnetic radiation: the closest quantum counterparts of classical states of electromagnetic field. In our approach to the quantum theory of electromagnetism, which starts with the quantum mechanics of photons, the quantum theory of the electromagnetic field can be obtained most easily by the procedure of second quantization. According to this procedure the photon wave functions appearing in the Fourier expansion (\ref{pwdec}) of the RS vector are upgraded to the operators: the annihilation and creation operators.
\begin{eqnarray}\label{ancr}
f_\pm(\bm k)\to a_\lambda(\bm k),\quad f_\pm^*(\bm k)\to a_\lambda^\dagger(\bm k).
\end{eqnarray}
These operators obey the commutation relations:
\begin{eqnarray}\label{crel}
\left[a_\lambda(\bm k),a_{\lambda'}^\dagger(\bm k')\right]=\delta_{\lambda\lambda'}k\delta^{(3)}(\bm k-\bm k').
\end{eqnarray}
The appearance of the factor $k$ in this formula is due to our (relativistic) normalization of the photon wave functions that is carried over to the annihilation and creation operators.

Afer second quantization, the RS vector (\ref{pwdec}) and its complex conjugate become the field operators built from annihilation and creation operators,
\numparts
\begin{eqnarray}
\fl\qquad{\hat{\bm F}}(\bm r,t)=\sqrt{\hbar c}\int\!\frac{d^3k}{(2\pi)^{3/2}}{\bm e}(\bm k)\left[a_+(\bm k)\rme^{\rmi(\bm k\cdot\bm r-\omega t)}+a^\dagger_-(\bm k)\rme^{-\rmi(\bm k\cdot\bm r-\omega t)}\right],\label{rsp}\\
\fl\qquad{\hat{\bm F}^\dagger}(\bm r,t)=\sqrt{\hbar c}\int\!\frac{d^3k}{(2\pi)^{3/2}}{\bm e}^*(\bm k)\left[a_-(\bm k)\rme^{\rmi(\bm k\cdot\bm r-\omega t)}+a^\dagger_+(\bm k)\rme^{-\rmi(\bm k\cdot\bm r-\omega t)}\right].\label{rsm}
\end{eqnarray}
\endnumparts
According to (\ref{rs}) the field operators of the electromagnetic field are the following linear combinations of ${\hat{\bm F}}$ and ${\hat{\bm F}^\dagger}$,
\begin{eqnarray}\label{db}
{\hat{\bm D}}
=\sqrt{\frac{\varepsilon_0}{2}}\left({\hat{\bm F}}+{\hat{\bm F}^\dagger}\right),
\quad{\hat{\bm B}}
=\frac{1}{\rmi}\sqrt{\frac{\mu_0}{2}}\left({\hat{\bm F}}-{\hat{\bm F}^\dagger}\right).
\end{eqnarray}
From these equations we obtain the decomposition of the electric and magnetic field into annihilation and creation operators.

In quantum field theory the photon wave functions still play the same role as in quantum mechanics of photons. Namely, they specify which photon states are created by the creation operators. In particular, the general one-photon state $|{\bm{\mathfrak f}}\rangle$ is obtained by acting on the vacuum state vector $|0\rangle$ with the creation operators $a^\dagger_\pm(\bm k)$ smeared with the functions $f_\pm(\bm k)$,
\begin{eqnarray}\label{cr}
a^\dagger({\bm{\mathfrak f}})=\sum_\lambda\int\!\frac{d^3k}{k}
f_\lambda(\bm{k})a_\lambda^\dagger(\bm{k}),\qquad |{\bm{\mathfrak f}}\rangle=a^\dagger({\bm{\mathfrak f}})|0\rangle.
\end{eqnarray}
One may recover the photon wave functions from the relations:
\begin{eqnarray}\label{hint}
f_+(\bm k)=\langle 0|a_+(\bm{k})|{\bm{\mathfrak f}}\rangle, \qquad f_-(\bm k)=\langle 0|a_-(\bm{k})|{\bm{\mathfrak f}}\rangle.
\end{eqnarray}

A coherent state $|{\bm{\mathfrak f}}_{\rm coh}\rangle$ of the electromagnetic field is associated with the photon wave function through the Glauber displacement operator $D({\bm{\mathfrak f}})$ \cite{rg},
\begin{eqnarray}\label{coh}
|{\bm{\mathfrak f}}_{\rm coh}\rangle=D({\bm{\mathfrak f}})|0\rangle
=\rme^{\sqrt{N}\left[a^\dagger({\bm{\mathfrak f}})-a({\bm{\mathfrak f}})\right]}|0\rangle,
\end{eqnarray}
where the operators $a^\dagger({\bm{\mathfrak f}})$ and $a({\bm{\mathfrak f}})$ are defined by the formula (\ref{cr}) and its Hermitian conjugate and $N$ is the average number of photons in the coherent state. The displacement operator $D({\bm{\mathfrak f}})$ shifts the creation and annihilation operators by $\sqrt{N}f^*$ and $\sqrt{N}f$ respectively,
\numparts
\begin{eqnarray}
D^\dagger({\bm{\mathfrak f}})a_\pm^\dagger(\bm k)D({\bm{\mathfrak f}})=a_\pm^\dagger(\bm k)+\sqrt{N}f_\pm^*(\bm k),\\
D^\dagger({\bm{\mathfrak f}})a_\pm(\bm k)D({\bm{\mathfrak f}})=a_\pm(\bm k)+\sqrt{N}f_\pm(\bm k).\label{shift}
\end{eqnarray}
\endnumparts
Therefore, the average value of the RS operator (\ref{rsp}) in a coherent state is described by the formula
\begin{eqnarray}\label{avcoh}
\qquad\qquad\fl{\bm F}(\bm r,t)&=\langle {\bm{\mathfrak f}}_{\rm coh}|\hat{\bm F}(\bm r,t)|{\bm{\mathfrak f}}_{\rm coh}\rangle\nonumber\\
&=\sqrt{\hbar cN}\int\!\frac{d^3k}{(2\pi)^{3/2}}{\bm e}(\bm k)\left[f_+(\bm k)\rme^{\rmi(\bm k\cdot\bm r-\omega t)}+f^*_-(\bm k)\rme^{-\rmi(\bm k\cdot\bm r-\omega t)}\right],
\end{eqnarray}
which differs from the corresponding formula (\ref{pwdec}) in quantum mechanics of a single photon only by the square root of the photon number in the coherent state. This formula gives the precise meaning, within the full quantized theory of the electromagnetic field, to the {\em one-to-one correspondence} between the one photon wave function $\bm{\mathfrak f}(\bm k)$ and the classical solution of the Maxwell equations: the real and the imaginary parts of ${\bm F}$. Namely, the classical solution is the average field in the coherent state which is made of photons described by the wave function $\bm{\mathfrak f}(\bm k)$. Using this correspondence, we may resolve the problem of the superposition of the waves with different helicities. Even though the photon wave functions with different helicities $f_+$ and $f_-$ cannot be simply added, the average field in the coherent state constructed from the two-component wave function ${\bm{\mathfrak f}}$ is the sum of the fields constructed separately from $f_+$ and $f_-$.

\section{Photon wave functions of some popular beams of radiation}

Exact solutions of the Maxwell equations describing the beams of radiation are so complex that they are rarely written in full detail. In contrast, the wave functions of photons in the momentum representation that make these beams are remarkably simple.

Even for the simplest beams---the Bessel beams---the electric and magnetic field vectors of the exact solutions of the Maxwell equations are fairly complicated. On the other hand, the states of photons comprising the beam are characterized simply by four quantum numbers. These are: the helicity $\pm\hbar$, the energy (or frequency $\omega$), the momentum in the $z$ direction (or the wave vector $q_z$), and the $z$-component of the total angular momentum $\hbar M$. The eigenfunctions in the momentum representation corresponding to these quantum numbers have the form (in cylindrical coordinates):
\begin{eqnarray}\label{bes}
f_{\rm B}(k_\perp,\phi,k_z)=\rme^{\rmi M\phi}\delta\left(\omega-c\sqrt{k_\perp^2+k_z^2}\right)
\delta(k_z-q_z).
\end{eqnarray}
The sign of helicity depends on whether we treat this function as $f_+(\bm k)$ or $f_-(\bm k)$. Upon the substitution of (\ref{bes}) into (\ref{avcoh}) and after performing the integrations, we obtain the exact solutions of Maxwell equations in the form of Bessel beams. However, the formula (\ref{bes}) is certainly much more informative than its counterpart in the position representation. At first glance, we can identify the relevant operators and their eigenvalues.

The exact solutions of Maxwell equations in the form of the Laguerre-Gauss beams \cite{bb0} are comprised of photons described by the wave functions that are the eigenfunctions of helicity $\pm\hbar$, the $z$-component of the total angular momentum $\hbar M$, and a peculiar operator reminiscent of the paraxial Schr\"odinger equation $\rmi\nabla_z+(c/4\Omega)(\nabla_x^2+\nabla_y^2)$ with eigenvalue $-\Omega/c$. Again the photon wave function in the momentum representation is relatively simple
\begin{eqnarray}\label{lg}
\fl\quad f_{\rm LG}(k_\perp,\phi,k_z)=\rme^{\rmi M\phi}\rme^{-\rmi (\Omega+c^2k_\perp^2/4\Omega)t}
\delta(k_z+ck_\perp^2/4\Omega-\Omega/c)k_\perp^{n+M/2}
\rme^{-l^2k_\perp^2/4},
\end{eqnarray}
while the solutions of the Maxwell equations would occupy half a page \cite{bb0}.

Finally, the exponential beams introduced in \cite{exp} have the following photon wave function in the momentum representation
\begin{eqnarray}\label{exp1}
\fl\qquad f_{\rm Exp}(k_\perp,\phi,k_z)=\rme^{\rmi M\phi}\rme^{-\rmi\sqrt{k_\perp^2+k_z^2}(t-\rmi\tau)}
\delta(k_z-q_z)k_\perp^M/\sqrt{k_\perp^2+k_z^2}.
\end{eqnarray}
The exponential beams share with Bessel beams the quantum numbers $M$ and $q_z$. They are not monochromatic but unlike Bessel beams they carry finite energy because they fall off exponentially in the direction perpendicular to the beam axis.

\section{Helicity, spin, and polarization}\label{sep}

Owing to the two-dimensional form of the photon wave function, there is no problem with identifying the positive (top) and the negative (bottom) helicity components. In the momentum representation the only condition that the helicity components must satisfy is the normalizability. In contrast, in the position representation the components of the wave function with opposite helicity have also different space-time structure. In particular, they obey different wave equations (\ref{curl}).

How does the notion of helicity carry over to the classical electromagnetic field? Can one split an arbitrary electromagnetic field into its positive and negative helicity parts? The answer is YES: The notion of helicity applies not only to photons but also to the classical electromagnetic field. The splitting can be done in two ways. We can expand the RS vector into plane waves (\ref{avcoh}) and separate the positive and negative frequency parts. Alternatively, we can do it even without the Fourier decomposition with the help of the projection operators $P_\pm=(1\pm\hat{\chi})/2$ where \cite{loc},
\begin{eqnarray}\label{chi}
\hat{\chi}{\bm F}(\bm r,t)=\frac{1}{2\pi^2}\int\!d^3r'\frac{1}{|\bm r-\bm r'|^2}{\bm\nabla}\times{\bm F}(\bm r',t).
\end{eqnarray}
The operator $\hat{\chi}$ is the helicity operator: it is equal to the projection of the angular momentum (\ref{gena1}) on the direction of the momentum (\ref{genc1}). The wave functions $\Psi_\pm$ are its eigenfunctions belonging to the eigenvalues $\pm 1$. Given an arbitrary solution of the Maxwell equations, described by the RS vector, we may split this field into the two helicity components by applying the projection operators,
\begin{eqnarray}\label{proj}
{\bm\Psi}_+(\bm r,t)= P_+{\bm F}(\bm r,t),\qquad
{\bm\Psi}_-(\bm r,t)= \left(P_-{\bm F}(\bm r,t)\right)^*.
\end{eqnarray}
Therefore, every classical solution of the Maxwell equations can be decomposed into two solutions ${\bm F}(\bm r,t)={\bm\Psi}_+(\bm r,t)+\left({\bm\Psi}_-(\bm r,t)\right)^*$, each corresponding to one sign of helicity. Note that the separation of the helicity components is a {\em nonlocal} operation; we must know the field in all space to determine its helicity components. This fact is also underscored in the quantum theory of the electromagnetic field. The part of the field operator $\hat{\bm F}$ built from the annihilation or creation operators of only one helicity is a nonlocal operator since its components do not commute at space-like separations \cite{loc1}.

The notion of spin of the photon is often used in the literature but when it comes to a precise definition we encounter a severe problem. It might seem that the spin is clearly identified as the second term of the total angular momentum (\ref{genc1}). However, it turns out that ${\bm s}$ cannot be treated as a stand-alone operator. The spin operators acting on the photon wave functions $\bm\Psi$ produce functions that do not belong to the physical space of divergence-free states; they contain an unphysical, longitudinal component. For example, ${\bm\nabla}\!\cdot\!s_z\bm\Psi=\rmi(\nabla_y\Psi_x-\nabla_x\Psi_y)$. The same problem appears in the first term of the total angular momentum (\ref{genc1}). This term looks like the standard quantum-mechanical orbital angular momentum but again it takes the photon wave functions out of the physical space ${\bm\nabla}\!\cdot\!(-\rmi x\partial_y+\rmi y\partial_x)\bm\Psi=-\rmi(\nabla_y\Psi_x-\nabla_x\Psi_y)$. It is only in the total angular momentum that the unwanted terms cancel out and the divergence-free condition is satisfied.

Even though the operators of the orbital angular momentum $-\rmi{\bm r}\times{\bm\nabla}$ and spin $s_i$ in the position representation are ill-defined, their expectation values calculated according to the formula (\ref{av}) are meaningful because the longitudinal component is canceled and we reproduce exactly the average values of both terms in (\ref{genc}),
\numparts
\begin{eqnarray}\label{avang}
\int\!d^3r\,{\bm\Phi}_{\pm}^*\left(-\rmi{\bm r}\times{\bm\nabla}\right){\bm\Phi}_{\pm}
=\int\frac{d^3k}{k}f_\pm^*\left(-\rmi{\bm k}\times{\bm D}\right)f_\pm,\\
\int\!d^3r\,{\bm\Phi}_{\pm}^*{\bm s}{\bm\Phi}_{\pm}=\pm\int\frac{d^3k}{k}f_\pm^*{\bm n}_{\bm k}f_\pm.
\end{eqnarray}
\endnumparts

It might seem that at least in the momentum representation the separation of the total angular momentum operator (\ref{genc}) into the orbital angular momentum part and the spin part is well defined, but the two terms have no acceptable eigenfunctions. For example, the $z$-component of the orbital part is $-\rmi\partial_\phi-\cos\theta$ and this operator has no single-valued eigenfunctions. This problem does not arise for the $z$-component of the total angular momentum since the troublesome term $\cos\theta$ is canceled by the contribution from the spin term and the operator $J_z$ has the eigenfunctions $\rme^{\rmi M\phi}$. Thus, the spin cannot be disentangled from the total angular momentum. Therefore, the spin of the photon cannot be interpreted as an additional degree of freedom; it cannot be separately manipulated. Moreover, its average direction is either parallel (for positive helicity states) or antiparallel (for negative helicity states) to the wave vector $\bm k$.

The true extra degree of freedom (in additional to the translational degrees of freedom) is the {\em polarization} made available due to the two-component form of the photon wave function. In the general sense, polarization is just this degree of freedom that offers us the choice to take {\em any combination} of the two helicity components to form a valid photon wave function or a classical electromagnetic field.

The standard tool in the description of polarization are the Stokes parameters \cite{spie}. These parameters can be introduced in the most general setting as the following bilinear combinations of the two helicity components:
\begin{eqnarray}\label{stokes}
S_0={\underline{\bm\Psi}}^\dagger{\underline{\bm\Psi}},\;
S_1={\underline{\bm\Psi}}^\dagger\rho_1{\underline{\bm\Psi}},\;
S_2={\underline{\bm\Psi}}^\dagger\rho_2{\underline{\bm\Psi}},\;
S_3={\underline{\bm\Psi}}^\dagger\rho_3{\underline{\bm\Psi}}.
\end{eqnarray}
The parameter $S_0$ describes the intensity of the wave and the parameter $S_3$ describes the preponderance of the positive helicity wave over the negative helicity wave. In classical optics the notion of polarization is applied only to some special optical beams, usually monochromatic plane waves, for which the Stokes parameters are constant. Also, they are usually defined in the linear polarization basis but this only amounts to a renaming of the parameters $S_i$. From Stokes parameters, we can construct a unit vector $\{S_1,S_2,S_3\}/S_0$ that gives a visual representation of the polarization states. Every polarization state is represented by a vector whose end lies on a sphere called the Poincar\'e sphere.

The three examples of photon wave functions  (\ref{bes}-\ref{exp1}) underscore the fact that the photon spin plays no role in the analysis of beams endowed with angular momentum: the quantum number $M$ refers to the total angular momentum, helicity is determined by the position of the wave function $f(\bm k)$ in the photon wave function $\mathfrak{f}(\bm k)$ and the polarization is determined by the composition of this two-component wave function.

In empty space the notion of polarization is not very interesting because both components propagate independently. It acquires physical significance when the decomposition into two helicity components changes during the time evolution due to the interaction with some medium, as shown in the next section.

\section{Propagation of photons in a medium}\label{med}

Quantum mechanics of photons moving in empty space is a mathematically rigorous theory but important physical problems appear when photons move in various media. Strictly speaking, photons moving in a medium are not the elementary particles but rather collective excitations of the electromagnetic field and the medium, like phonons, polarons, excitons, etc. However, the propagation of light in a static medium can be fully described in terms of the quantum mechanics of (effective) photons. It is not so in a medium, whose characteristics vary with time because then a single photon description is not sufficient; photons are produced and absorbed by the parametric mechanism (dynamical Casimir effect).

The evolution equations for the photon wave function in a medium follow from the Maxwell equations. They are obtained by taking the positive frequency parts of the coupled equations for the vectors $\bm F$ and $\bm F^*$,
\begin{eqnarray}\label{eqm}
\fl\qquad\rmi\partial_t{\bm\Psi}_+=v(\bm r)\left[{\bm\nabla}\times{\bm\Psi}_++{\bm\nabla}\ln\sqrt{v(\bm r)}\times{\bm\Psi}_++{\bm\nabla}\ln\sqrt{h(\bm r)}\times{\bm\Psi}_-\right],\\
\fl\qquad\rmi\partial_t{\bm\Psi}_-=-v(\bm r)\left[{\bm\nabla}\times{\bm\Psi}_-+{\bm\nabla}\ln\sqrt{v(\bm r)}\times{\bm\Psi}_-+{\bm\nabla}\ln\sqrt{h(\bm r)}\times{\bm\Psi}_+\right],
\end{eqnarray}
where $v(\bm r)=1/\sqrt{\varepsilon(\bm r)\mu(\bm r)}$ is the local velocity of light and $h(\bm r)=\sqrt{\mu(\bm r)/\varepsilon(\bm r)}$ is the local impedance of the medium. There are some errors in signs in these equations in Ref.~\cite{pwf}. It is worth noticing that it is only the impedance (but not the velocity) that couples the two helicities. In particular, in curved space the impedance is constant and the two helicity components propagate independently; i.e. the polarization is constant.

\section{Conclusions}

The reliance on the solid foundation, namely on the theory of representations of the Poincar\'e group developed by Wigner, enabled us to formulate the quantum mechanics of photons. In this formulation, the notions of helicity, angular momentum, spin, and polarization acquire a precise meaning and we are able to rectify many misinterpretations found in the literature. According to the theory of representations, the photon helicity is the most fundamental property; photons with opposite helicities are separate relativistic objects. Further, we have shown that the polarization and not the spin is the true degree of freedom. The spin appears only as a part of the total angular momentum and it cannot be separated from the orbital angular momentum.
Quantum mechanics of photons is further extended, with the use of the second quantization, to the full quantum theory of the electromagnetic field. Within the framework developed in this work, the notion of the wave-particle duality acquires a precise mathematical meaning.

\ack
This research was financed by the Polish National Science Center Grant No. 2012/07/B/ST1/03347.

\end{document}